\documentclass[%
preprint,
 amsmath,amssymb,
]{revtex4-1}
\usepackage{graphicx}
\usepackage{bm}
\usepackage{lineno}

\begin{document}
\title{
Efficient thermo-spin conversion in van der Waals ferromagnet FeGaTe
}

\author{Shuhan Liu}
\affiliation{Department of Physics, Kyushu University, 744 Motooka, Fukuoka, 819-0395, Japan}

\author{Shaojie Hu}
\email[]{hu.shaojie@phys.kyushu-u.ac.jp} 
\affiliation{Department of Physics, Kyushu University, 744 Motooka, Fukuoka, 819-0395, Japan}

\author{Xiaomin Cui}
\affiliation{Department of Physics, Kyushu University, 744 Motooka, Fukuoka, 819-0395, Japan}

\author{Takashi Kimura}
\email[]{t-kimu@phys.kyushu-u.ac.jp}
\affiliation{Department of Physics, Kyushu University, 744 Motooka, Fukuoka, 819-0395, Japan}

\date{\today}

\begin{abstract}

Recent discovery of 2D van der Waals (vdW) magnetic materials has spurred progress in developing advanced spintronic devices. A central challenge lies in enhancing the spin-conversion efficiency for building spin-logic or spin-memory devices. We systematically investigated the anomalous Hall effect and anomalous Nernst effect in above-room-temperature van der Waals ferromagnet FeGaTe with perpendicular anisotropy, uncovering significant spin-conversion effects. The anomalous Hall effect demonstrated an efficient electric spin-charge conversion, with a notable spin Hall angle of 6 $\%$ - 10.38 $\%$. The temperature-dependent behavior of the anomalous Nernst voltage primarily results from the thermo-spin conversion. Uniquely, we have experimentally achieved thermo-spin polarization values of over 690 $\%$  at room temperature and extremely large of 4690 $\%$ at about 93 K. This study illuminates the potential of vdW ferromagnets in advancing efficient spin conversion devices.\\~\\


\end{abstract}

\maketitle
\section{Introduction}
The recent elucidation of two-dimensional (2D) van der Waals (vdW) magnetic materials heralds a wealth of prospects for the advancement of future spintronic devices and the investigation of spin-related physical phenomena.\cite{lee2016ising,gong2017discovery,huang2017layer,gibertini2019magnetic} These phenomena encompass quantum spin phases transition, topological spin excitations, and spin-conversion. \cite{kimura2007room,otaniSpinConversionNanoscale2017,liMagneticFieldInducedQuantumPhase2020,chenMagneticFieldEffect2021} Existing literatures intimate that magnetism in 2D vdW magnets and their multifunctional heterostructures can be considerably modulated by factors such as electric gating, strain, heat, and light.\cite{deng2018gate,wang2023interfacial} 
A profound understanding of spin-conversion modulation mechanisms within 2D vdW magnetic systems is expected to invigorate forthcoming research, paving the way for pioneering applications in spintronics, spin-caloritronics, and quantum spin computation.
The effective generation, injection, and control of spin current (a flow of magnetic angular momentum) remain pivotal for the development of future efficient spin-conversion devices.
The discovery of the spin Seebeck effect, generating thermo-spin current by the temperature gradient in ferromagnet, provides a novel approach for the spin conversion by using heat current.\cite{uchida2008observation}
Two-dimensional vdW ferromagnetic insulators such as CrI$_3$\cite{marfoua2020giant}, CrPbTe$_3$\cite{marfoua2021spin} and MnCl$_3$\cite{li2021spin} are forecasted to demonstrate efficient spin Seebeck effects. 
In ferromagnets, the anomalous Nernst effect (ANE), a heat current generates a transverse voltage, is considered as the coupling of thermo-spin and spin-charge conversion.\cite{fang2016scaling}
A substantial anomalous Nernst effect in van der Waals ferromagnet $\rm Fe_3GeTe_2$ is attributed to its topological nature and the large Berry curvature near the Fermi level.\cite{xu2019large,yang2022sign} 
The maximum anomalous Nernst angle ($S_{xy}/S_{xx}$) is observed to be 3 at low temperature\cite{fang2019observation}, which significantly surpasses the performance of conventional room temperature ferromagnets, such as ferromagnetic binary compounds (Fe$_3$Al, Fe$_3$Ga, Fe$_3$Pt)\cite{sakai2020iron,li2023large}, multilayer films\cite{fang2016scaling,hasegawa2015material,uchida2015enhancement} and Heusler composites\cite{sakai2018giant}.  

Despite the recent advancements, a notable limitation persists in the realm of the aforementioned two-dimensional (2D) magnets: their Curie temperature remains below room temperature, which poses significant challenges to practical application. To facilitate the realization of high-quality and practical devices for such applications, there is a compelling need for robust 2D magnets exhibiting ordering temperatures above room temperature. 
Recent endeavors have focused on the exploration of 2D van der Waals magnetic materials with temperatures exceeding room temperature. As an illustration, Fe-rich van der Waals ferromagnets, $\rm Fe_nGeTe_2$ (n=4-5), boast not only substantial perpendicular magnetic anisotropy but also a $\rm T_c$ that is progressively nearing room temperature.\cite{may2019ferromagnetism,seo2020nearly,zhang2020itinerant,mondal2021critical,wang2023interfacial}  Further enhancements in the Curie temperature, up to 400 K - 530 K, can be achieved through meticulous interfacial engineering. \cite{li2022magnetic,wang2020above,wang2023interfacial} 
Additionally, cobalt-substituted  Fe$_5$GeTe$_2$ \cite{chen2023above} also demonstrates a $\rm T_c$ that surpasses room temperature. This discovery underpins the potential of these 2D magnets for various applications, highlighting the importance of continued research in this field.
Recently, an innovative addition to the realm of two-dimensional van der Waals ferromagnetic conductors, Fe$_3$GaTe$_2$, has been fortuitously identified.\cite{zhangAboveroomtemperatureStrongIntrinsic2022} Its Curie temperature is reported to be in the range of 350K to 380K, marking a significant milestone as the highest recorded for known 2D magnets. It demonstrates noteworthy characteristics such as an elevated saturation magnetic moment, steadfast perpendicular magnetic anisotropy, and a substantial anomalous Hall angle ($\theta_{AH}$) at room temperature. These findings suggest that Fe$_3$GaTe$_2$ possesses an augmented spin-charge conversion, thereby offering a promising platform for exploring the efficient thermo-spin conversion. In this study, we aim to scrutinize the efficiency of thermal-spin conversion in this material by harnessing the Anomalous Nernst Effect (ANE). It is to be noted that the nontrivial temperature-dependent behavior of the ANE can be predominantly attributed to the generation of thermo-spin current. 

\section{Experimental results}

The single crystals of FeGaTe were produced by utilizing the chemical vapour transport method with transport agent iodine ($\rm I_2$). First, the high-purity stoichiometric quantity powders of Fe, Ga, Te, and $\rm I_2$ were sealed in a quartz tube. The quartz tube was set in a two-zone furnace for 170 hours to grow the bulk FeGaTe crystals with the size of $\rm 7mm \times 8 mm$. 
The high-resolution high-angle annular dark-field scanning transmission electron microscopy image reveals the high-quality crystal structure of the FeGaTe in Fig. 1(a). Cross-sectional scanning transmission electron microscopy (STEM) image with lower magnification shows all the layers are regularly arranged with clear vdw gap. And atomic resolution energy-dispersive X-ray spectroscopy (EDS) mapping in the insert of Fig.1(b) shows that all Fe, Ga, and Te elements are uniformly distributed in the film. Together, these results prove that the high-quality single-crystalline FeGaTe has been obtained. Chemical analysis using EDS confirms the stoichiometry of $\rm Fe_{2.78}Ga_{0.88}Te_{2.34}$ in Fig.1(b). 
The room temperature magnetic hysteresis loops (M-H loop) of bulk FeGaTe were measured by using the vibrating sample magnetometer (VSM). Figure.1(c) shows the typical magnetic hysteresis loops under the out-of-plane  (red) and in-plane (blue) magnetic field directions at 295 K. The lower coercivity and larger remagnetization for the out-of-plane curve indicate a large perpendicular magnetic anisotropy property of such crystal. Then, the field-cooled (FC) thermal magnetization curve was measured using a superconducting quantum interference device (SQUID).  
Figure.1(d) shows the M-T curve with a 3600 Oe magnetic field applied along the c-axis of the bulk crystal with the initial cooling from 300 K. The experimental data of M-T is well fitted by the function $M=M_0(1-\frac{T}{T_C})^\beta$ as shown in the solid line\cite{evans2014atomistic}. The values of $\beta$ and Curie temperature $T_c$ were fitted to be 0.3 and 365 K, respectively, where $T_C$ obviously exceeds the room temperature. 

To evaluate the anomalous Hall effect and anomalous Nernst effect with the same devices, we fabricated the Hall bar structure devices with the zigzag heater for the Fe$_3$GaTe$_2$ thin flakes on $\rm SiO_2/Si$ substrate. The scanning electron microscopy (SEM) image of the device is shown in Fig.2(a).
Here, the thin flakes were mechanically exfoliated from the bulk crystals and transferred onto SiO$_2$/Si substrate by scotch tape. 
All the electrodes were fabricated by using electron beam lithography (EBL) and a wet lift-off process. 
The first EBL process was used to fabricate the 300 nm Cu electrodes deposited in the ultra-high vacuum Joule evaporator. Before the Cu deposition, we used the low acceleration argon ion beam to clean the contact surface to obtain good adhesion. 
The second EBL process was used to fabricate the zigzag heaters with 200 nm Cu. Before depositing the Cu, 50nm SiO$_2$  was deposited by magnetron sputtering for electrical isolation between FeGaTe and heaters. Both anomalous Hall and Nernst effects were evaluated in a cryostat with a temperature-control system. The anomalous Hall and Nernst voltages were detected by a lock-in amplifier with first- and second-harmonic technology.\cite{hu2016first} 

The experimental setup for detecting anomalous Hall effect (AHE) is illustrated in Fig. 2(b), where we applied a small A.C. current of 173 Hz flowing along x-axis ($I_x$), along with an out-of-plane magnetic field. The transverse voltage ($V_{xy}$) and longitudinal ($V_{xx}$) voltage were measured by extracting the first-harmonic signal of lock-in amplifier. 
The temperature dependence of longitudinal resistance ($R_{xx}$) and conductivity ($\sigma_{xx}$) is shown in Fig. 2(c). The observed decreasing resistance with reducing temperature suggests a typical metallic behavior of the crystal. There is a minimum value of $R_{xx}$ at 30 K, followed by an upturn with the further decrease of temperature. This low-temperature behaviour may be induced by the Kondo scattering\cite{beal-monodKondoResistivityDue1969}. 
Then, we conducted the anomalous Hall voltage measurement with the magnetic field sweeping from -4k Oe to 4k Oe at various fixed temperatures. 
Figure 2(d) shows the field dependence of transverse resistance ($R_{xy}=V_{xy}/I_x$), and the anomalous Hall resistance is then defined as $R_{AHE}=\frac{\Delta R_{xy}}{2}$. Generally, the value of $R_{AHE}$ increases with the reduction of temperature. 
Notably, the rectangle-shaped hysteresis loops demonstrate the single domain state and robust perpendicular magnetic anisotropy (PMA) of Fe at lower temperatures. 
However, the magnetization became unsaturated at zero field with the further rise of temperature, as depicted by the red solid line (representing 315 K) in Fig. 2(d). 
A small jump or dip is observed for $R_{xy}$ before reaching zero fields. The R-H curve also transforms into a narrow-waist shape, which can be attributed to the emergence of vortex-like, skyrmion, or multi-domain wall textures before sweeping the field to zero\cite{guslienko2001Field,guslienko2001Magnetization,tan2018hard,xu2019large,gao2022Manipulation}. 
The similar feature has also been obtained in 2D van der Waals layered $\rm Fe_3GeTe_2$ due to the bulk/interface DMI effect, which is strongly related to the thickness of the flakes.\cite{tan2018hard,park2021neel,pengTunableNeelBloch2021} 
Thinner flakes exhibit a hysteresis shape closer to a square shape at room temperature, as shown in Fig. S3 of the Supporting Information. This suggests that thinner flakes will possess higher perpendicular anisotropy.
For a comprehensive understanding of AHE, we also investigated the temperature dependence of $R_{AHE}$ by cooling the sample from 315 K to 5 K and applying an out-of-plane magnetic field. This aims to guarantee the measured value of $R_{xy}$ corresponds to the state of magnetization saturation. 
The magnitude of the field of cooling (FC) was set to be $\lvert H \rvert$ =3600 Oe, which far exceeded the value of coercive field ($H_C$) at high temperatures, ensuring the magnetization remained in a saturated state. Meanwhile, considering the metallic nature of FeGaTe, the possible contribution from normal Hall effect caused by the external magnetic field can be neglected. 
In order to eliminate the longitudinal resistivity contribution due to contact electrode misalignment, the measurements were conducted under a positive field ($+H$) and a negative field ($-H$), respectively, where we defined $R_{AHE}=\frac{\Delta R_{xy}}{2}=\frac{R_{xy}^{+H}-R_{xy}^{-H}}{2}$. 
The transverse conductivity was calculated by $\sigma_{xy}=-\frac{\rho_{xy}}{\rho_{xx}^2+\rho_{xy}^2}\approx\ -\frac{\rho_{xy}}{\rho_{xx}^2}$, and its corresponding temperature dependence is shown in Fig. 2(e), which increases with the temperature decrease. 
The value of $\sigma_{xy}$ reaches 182 $\Omega^{-1}$cm$^{-1}$ at low temperature, which is of the same order as the intrinsic contribution from Berry curvature $\sigma_{xy}^{in}=\frac{e^2}{ha_c}\approx$ 238 $\Omega^{-1}$cm$^{-1}$, where $e$ is the electronic charge, $h$ is the Planck constant, and $a_c$ is the crystal parameter of 16.2290 \AA\cite{onoda2006intrinsic,xiao2010berry}.  

Based on the spin-charge conversion model illustrated in Fig.3(a), the detected transverse voltage of AHE can be simply considered as the result of the sequential interplay between electric-driven spin current and inverse spin Hall effect (ISHE).\cite{takahashi2008Spin} 
The spin current ($J_S$) is first generated due to the charge current flowing through the ferromagnet being polarized. And then the inverse spin Hall effect, which acts like a fictional magnetic field to cause electrons with different spins to scatter along different trajectories, results in electrical potential difference at the transverse ends of FeGaTe. Considering the inherent connection between AHE and ISHE, the quantitative relationship between anomalous Hall angle ($\theta_{AH}$) and inverse spin Hall angle ($\theta_{ISH}$, equivalent to spin Hall angle $\theta_{SH}$) can be assumed  as follows with the limitation of  an isotropic
spin polarization (P) and zero spin Hall angle polarization($P_{\theta}=(\theta_{\uparrow}-\theta_{\downarrow})/(\theta_{\uparrow}+\theta_{\downarrow})$):\cite{fang2016scaling,omori2019Relation} 
\begin{equation}
    \theta_{ISH}=\theta_{SH}=\frac{\theta_{AH}}{P}=\frac{\sigma_{xy}}{\sigma_{xx}}\frac{1}{P}
\end{equation}
In our analysis, $\theta_{AH}$ is experimentally determined by $\theta_{AH}=\frac{\sigma_{xy}}{\sigma_{xx}}$, as Fig. 3(b) exhibits, which is about 2.6 $\%$ at room temperature and reaches its maximum of 6.8 $\%$ at 5 K. While P is assumed to be proportional to the magnetization (M) based on the conclusion of previous reports, which can be mathematically described as $P=P_0(1-\frac{T}{T_C})^\beta$ \cite{mukhopadhyayTemperatureDependenceTransport2007,mukhopadhyayPointContactAndreev2008, kambojTemperatureDependentTransport2019,zhu2022large}. $P_0$ is the spin polarization at 0 K, and its value is estimated to be about 67.68 \% based on the previous report\cite{zhu2022large}, and the value of $\beta=0.3$ is used based on the simulation of the M-T curve in Fig.1 (d). The obtained temperature dependence of P is presented in Fig3. (c). Consequently, the derived $\theta_{SH}$ at varying temperatures is shown in Fig.3 (d), in the range of 6 $\%$ - 10.38 $\%$, which is much larger than that observed in some heavy metals and their alloys\cite{sinova2015spin}. As indicated by the results, FeGaTe is an effective electric spin-charge conversion material.

The thermo-spin properties were then examined via anomalous Nernst effect using a homemade measurement system equipped with three lock-in amplifiers, as shown in Fig. 4(a). 
Here, one amplifier was used to detect the second-harmonic signal of the transverse voltage $V_{xy}^{2\omega}$, and the other two measured the heater resistance simultaneously to characterize the local temperature increase. 
In our experimental configuration, the heat was mainly generated by the Joule heating of the left z-shaped heater (red), while the right strip (blue) merely served as a temperature reference of the cold side. An A.C. heating current ranging from 3mA to 10mA was applied. 
The obtained field dependence of $V_{xy}^{2\omega}$ is displayed in Fig. 4(b). Similarly, we give the definition of anomalous Nernst voltage $\Delta V_{xy}^{2\omega}=\frac{V_{xy}^{2\omega}(+H)-V_{xy}^{2\omega}(-H)}{2}$. 
Figure 4(c) reveals a clear parabolic relationship between $\Delta V_{xy}^{2\omega}$ and heating current $I_h$, attesting to the thermal origin of the detected signal due to Joule heating. 

Fig. 5(a) presents the temperature-dependent ANE curves at a heating current of 10 mA. 
As temperature decreases, the $V_{xy}^{2\omega}-H$ curves tend to firstly decrease and then increase, where a sign reversal of $\Delta V_{xy}^{2\omega}$ surprisingly occurs between 242 K and 217 K. This behavior is significantly different from that of AHE, which maintains positive at all temperatures. 
To further explore this unexpected phenomenon, we next measured the temperature dependence of ANE voltage. 
The basic methods of measurement mirror those used for AHE, involving the cooling of the sample within the base temperature range from 315 K to 5 K under positive and negative magnetic fields, respectively. The ANE voltage is then determined as the half-difference of two $ V_{xy}^{2\omega}$ values measured under a cooling field of 3600 Oe applied in positive and negative directions.
As shown in Fig. 5(b), $\Delta V_{xy}^{2\omega}$ first decreases and then increases within the whole cooling process, where the sign of $\Delta V_{xy}^{2\omega}$ is reversed from positive to negative at about 223K. After that, the absolute value of ANE signal continues to increase until the peak of 1.68 $\rm \mu$V at about 115 K. Such nontrivial temperature dependence behavior is quite different from that of the anomalous Hall effect. 
To understand the origin, the transverse and longitudinal thermoelectric coefficients are then calculated as $S_{xy}=E_y/\nabla T_x$ and $S_{xx}=E_x/\nabla T_x$, respectively. 
The accurate ascertainment of temperature gradient values, consistent with the actual experimental conditions, is critical to data analysis.
Despite the measurements of heater resistance indicating local temperature increases, the derived temperature difference of the two heat sensors, $\Delta T=T_{hot}-T_{cold}$, falls short in accurately reflecting the genuine temperature conditions of FeGaTe flake.
This is due to the isolation layer of SiO$_2$ between FeGaTe and heaters exhibiting poor thermal conductivity, and thus it inhibits the efficient heat transfer from heater to FeGaTe. As a result, directly calculating $\nabla T_x$ using $\Delta T$ tends to overestimate $\nabla T_x$. To resolve this issue, we combined the experimental results with COMSOL simulations to obtain a more accurate value of $\nabla T_x$.
 (See Supporting Information for simulation details such as $\nabla T_x$ at varying base temperatures). 
The temperature dependence of $S_{xy}$ and $S_{xx}$ is presented in Fig. 5(c). 
The $S_{xy}$ reaches the minimum value of the -453 nV/K at the temperature of (130.9$\pm$3.1) K, which is slightly higher than that of $V_{xy}$ because of the modification of temperature gradient.
The sign reverse temperature of $S_{xy}$ is still the same as $V_{xy}$.  
We attribute the unique behavior of the $S_{xy}$ with temperature variation to the efficient thermo-spin current generation. 
The model of thermo-spin current is established for a more thorough understanding shown in Fig.6(a). Here, the generation of ANE voltage is considered as a two-step process. First, the thermo-spin current is generated in FeGaTe by temperature gradient due to the spin-dependent Seebeck effect.\cite{johnson1987thermodynamic,slachter2010thermally} The thermo-spin current could be expressed as $J_s=-\sigma_{xx}(1-P^2)P_sS_{xx}\nabla T_x/2$, where thermo-spin polarization $P_s=(S_{\uparrow}-S_{\downarrow})/S_{xx}=S_s/S_{xx}$ is a crucial parameter for thermo-spin current generation. Then, electrons of the spin current with different spins moving along different trajectories lead to the occurrence of transverse voltage by ISHE. Therefore, the non-zero thermo-spin current is a prerequisite for ISHE to generate transverse charge accumulation, and thus the transverse voltage. Consequently, the Seebeck coefficients of up-spin ($S_{\uparrow}$), down-spin ($S_{\downarrow}$), and spin-dependent Seebeck coefficient ($S_s$) can be derived as follows: \cite{slachter2010thermally,hu2014efficient,fang2016scaling}:

\begin{equation}
S_{\uparrow}=\frac{PS_{xy}\rho_{xx}}{(1-P)\rho_{xy}}+S_{xx}
\end{equation}
\begin{equation}
S_{\downarrow}=-\frac{PS_{xy}\rho_{xx}}{(1+P)\rho_{xy}}+S_{xx}
\end{equation}
\begin{equation}
S_s=S_{\uparrow}-S_{\downarrow}=\frac{2PS_{xy}\rho_{xx}}{(1-P^2)\rho_{xy}}
\end{equation}

Then, we plotted the temperature dependence of $S_{\uparrow}$, $S_{\downarrow}$, and $S_s$  based on the experimental data, as shown in Fig. 6(b). 
Note that $S_{\uparrow}$ and $S_{\downarrow}$ exhibit opposite signs, indicating up-spins and down-spins are moving in opposite directions under the same temperature gradient. This offers a reasonable explanation for the relatively small value of $S_{xx}$, which can be described as $S_{xx}=\frac{S_{\uparrow}\sigma_{\uparrow}+S_{\downarrow}\sigma_{\downarrow}}{\sigma_{\uparrow}+\sigma_{\downarrow}}$, of which the contributions from up-spin and down-spin cancel out each other\cite{hu2014efficient,hu2018substantial}. With the further decrease of temperature, the sign of the majority spin is reversed, indicating an opposite contribution, and thus leads to a similar sign change in $S_{xy}$. We also estimate the thermo-spin polarization $P_s=S_s/S_{xx}$ as a function of temperature, shown in Fig.6 (c). As can be seen, the absolute value of thermo-spin polarization can be extremely large of 4690 $\%$ at about 93 K, and it exceeds 690$\%$ at room temperature, which is the first time to experimentally obtain such giant thermo-spin polarization.  (See Supporting Information S5 for comparison of $P_S$ in different materials).

In conclusion, our comprehensive study on FeGaTe single crystal flakes reveals prominent spin-conversion efficiency via the anomalous Hall effect and anomalous Nernst effect. The observed spin Hall angle ranges between 6 $\%$ - 10.38 $\%$, affirming the competent electric spin-charge conversion. The distinctive temperature-dependent behavior of the anomalous Nernst voltage is predominantly attributed to thermo-spin conversion. Experimental results showcase impressive thermo-spin polarization exceeding 690 $\%$ near room temperature, with an exceptional value reaching 4690 $\%$ around 93 K. This research not only elucidates the coupling among charge, spin, and heat but also offers fresh perspectives on optimizing spin conversion in vdW ferromagnets.

\begin{acknowledgments}
This work is partially supported by National JSPS Program for Grant-in-Aid for Scientific Research (S)(21H05021), and Challenging Explor- atory Research (17H06227) and JST CREST (JPMJCR18J1).
The authors thank the Ultramicroscopy Research Center for TEM sample preparation by using focused ion beam (FIB). The high-resolution STEM observation was supported by “Advanced Research Infrastructure for Materials and Nanotechnology in Japan (ARIM)” of the Ministry of Education, Culture, Sports, Science and Technology (MEXT).
\end{acknowledgments}

\newpage 

\bibliographystyle{unsrt}
\bibliography{FGT}

\newpage
\begin{figure}[h]
    \centering 
    \includegraphics[width=6.3in]{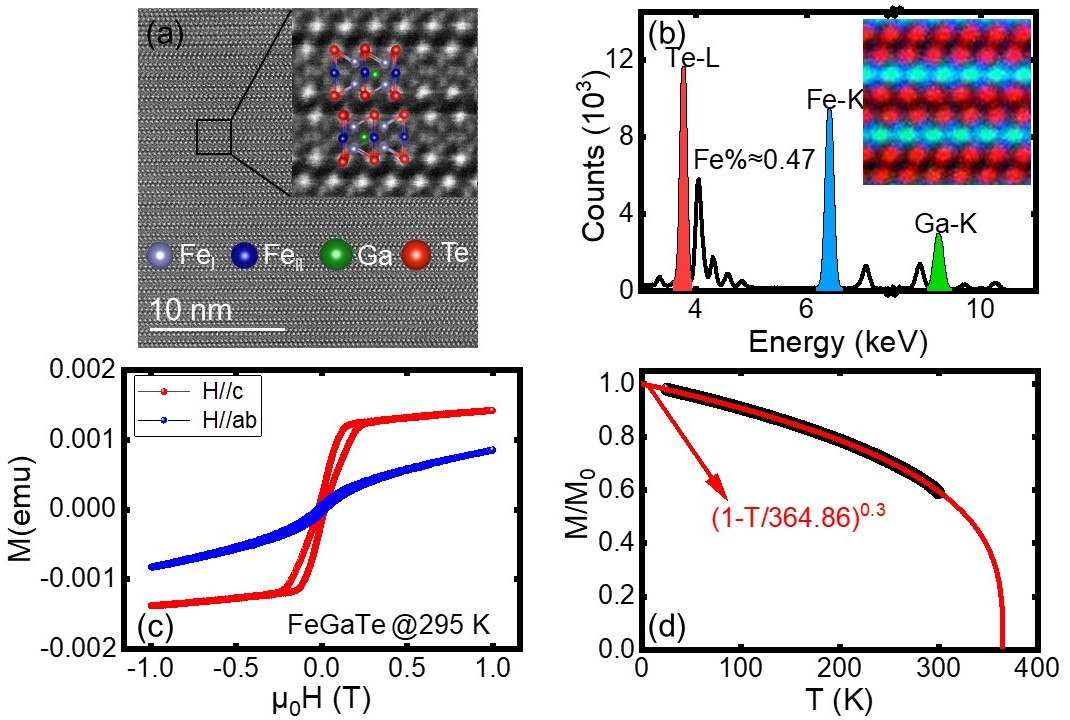}
    \caption{(a) The cross-sectional STEM image and crystal structure of FeGaTe. (b) Energy dispersive X-ray spectrum of FeGaTe flake. Inset shows atomic resolution EDS mappings optimized by the weak Wiener filter. The blue, green, and red spheres represent Fe, Ga, Te atoms, respectively. (c) M-H curves of bulk FeGaTe under out-of-plane (red) and in-plane (blue) magnetic fields at room temperature. (d) Temperature dependence of normalized spontaneous magnetization for bulk FeGaTe with cooling field 3600 Oe. The red solid line represents the data fitted through power law $M=M_0(1-\frac{T}{T_C})^\beta$.}
    \label{1}
\end{figure}

\begin{figure}[h]
    \centering 
    \includegraphics[width=6.3in]{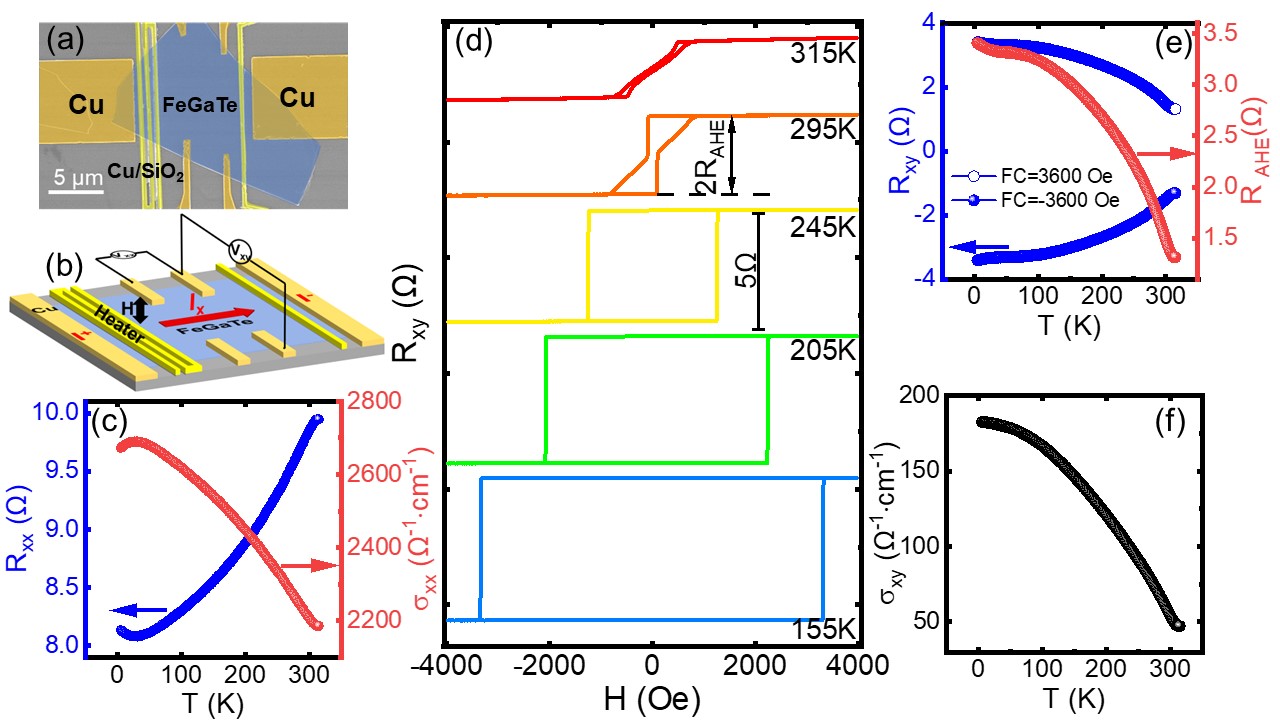}
    \caption{(a) The scanning electron microscope image of the device, where the blue, brighter yellow, and darker yellow areas represent the FeGaTe flake, heaters, and Cu electrodes, respectively. (b) Schematic of the AHE measurement setup. (c) The temperature dependence of longitudinal resistance $R_{xx}$ and conductivity $\sigma_{xx}$. 
    (d) The AHE signals obtained by measuring $R_{xy}$ with sweeping out-off plane magnetic field for various temperatures. (e) The temperature dependence of transverse resistance $R_{xy}$ under positive and negative magnetic cooling fields, and the derived AHE resistance $R_{AHE}$. (f) The temperature dependence of transverse conductivity $\sigma_{xy}$.}
    \label{2}
\end{figure}

\begin{figure}[h]
    \centering 
    \includegraphics[width=6.3in]{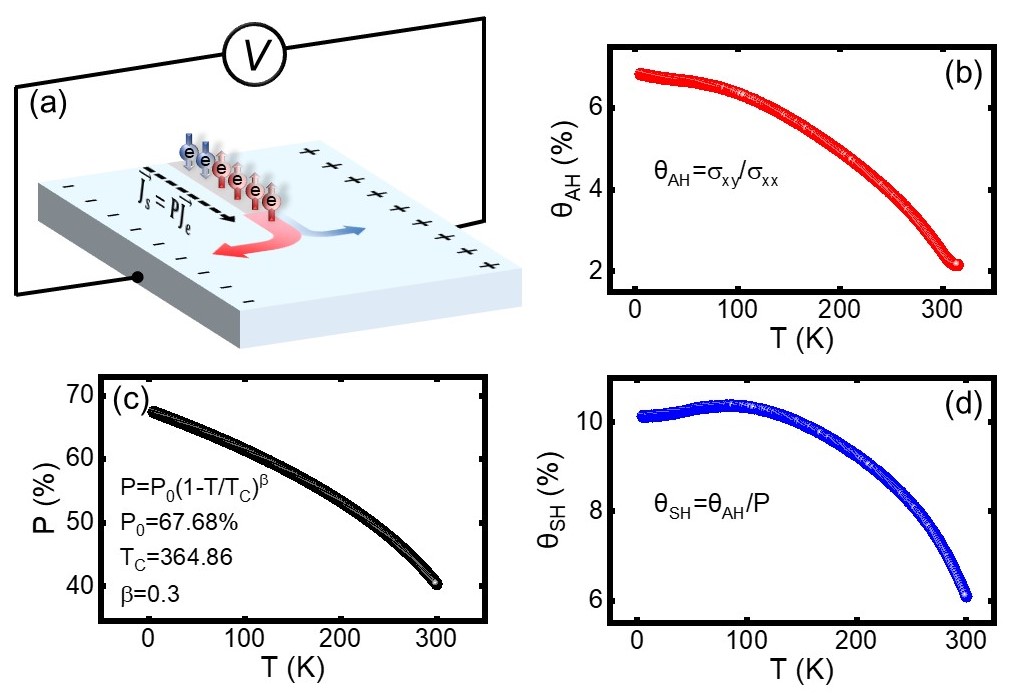}
    \caption{The illustration of AHE and ISHE in ferromagnet. The black dashed arrow represents the direction of electron flow ($\vec{J_e}$) and spin current ($\vec{J_s}$). Relative to the incident $\vec{J_e}$, transverse accumulations of spin and charge are detected as the SHE and AHE, respectively. (b-d) The temperature dependence of anomalous Hall angle $\theta_{AH}$, spin polarization $P$, and spin Hall angle $\theta_{SH}$ of FeGaTe. $P$  is obtained by $P=P_0(1-\frac{T}{T_C})^\beta$, where $P_0\ \approx \ 0.68$, $T_C=364.86$, and $\beta\ =\ 0.3$.}
    \label{3}
\end{figure}

\begin{figure}[h]
    \centering 
    \includegraphics[width=6.3in]{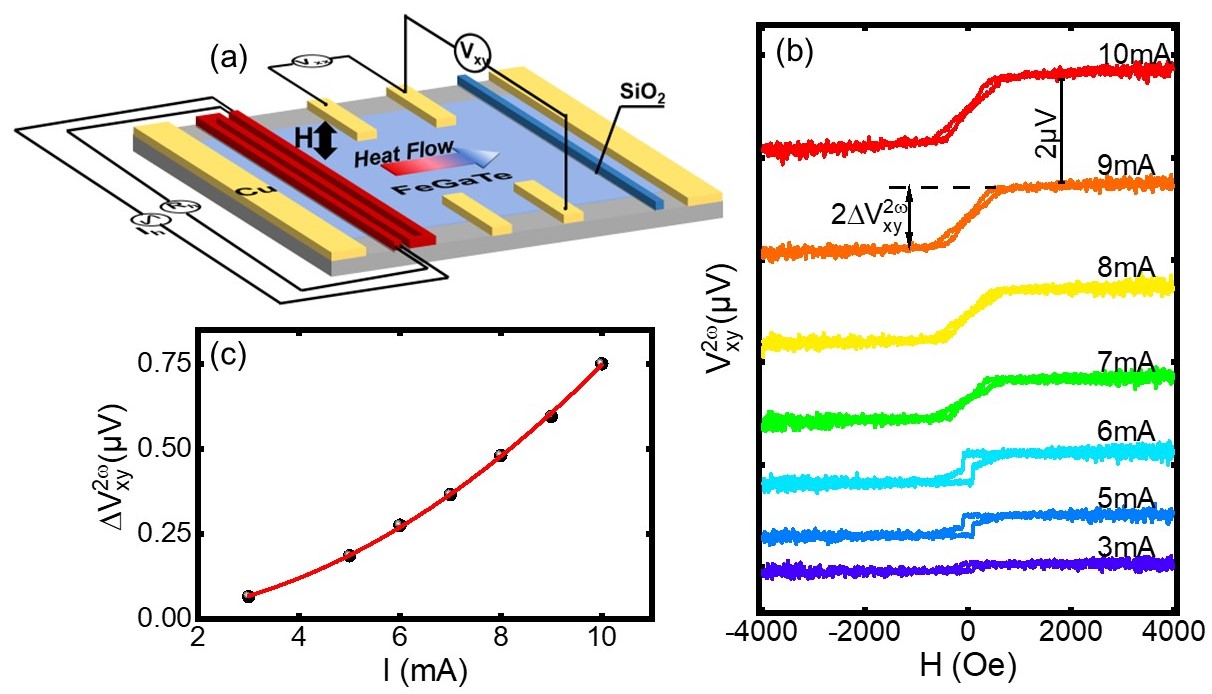}
    \caption{(a) Schematic of the ANE measurement setup.  (b) ANE signals obtained under varying heating currents with sweeping out-of-plane magnetic field at base temperature of 295 K. (c) The current dependence of ANE  voltage ($\Delta V_{xy}^{2\omega}$), which is well fitted by a typical parabolic relation.}
    \label{4}
\end{figure}

\begin{figure}[h]
    \centering 
    \includegraphics[width=6.3in]{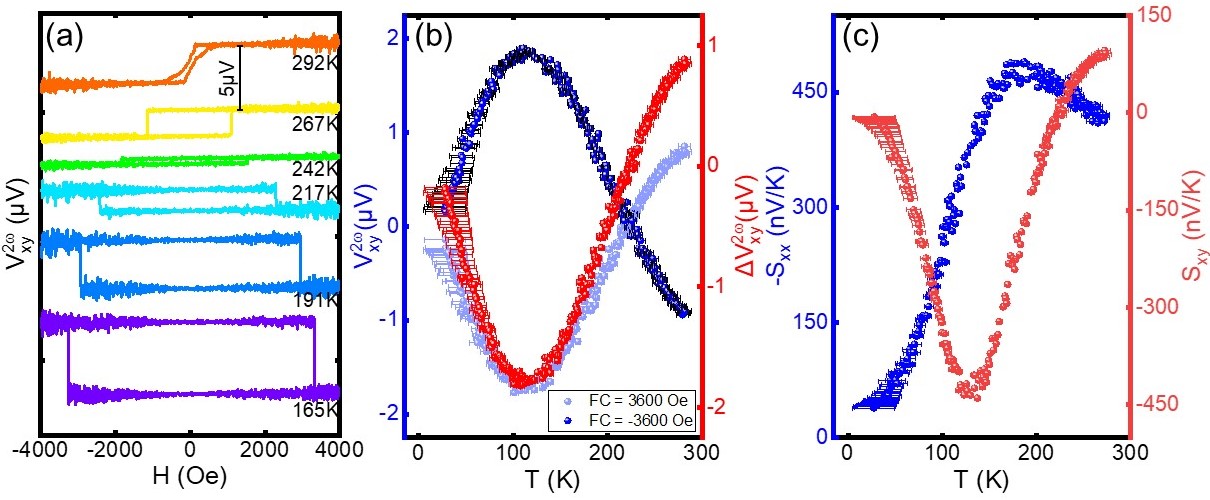}
    \caption{(a) ANE signals obtained by out-of-plane magnetic field sweeping under 10 mA bias current for various temperatures. (b) The temperature dependence of transverse voltage under positive and negative magnetic fields, and the derived ANE voltage. (c) The temperature dependence of $S_{xy}$ (blue) and $S_{xx}$ (red).}
    \label{5}
\end{figure}

\begin{figure}[h]
    \centering 
    \includegraphics[width=6.3in]{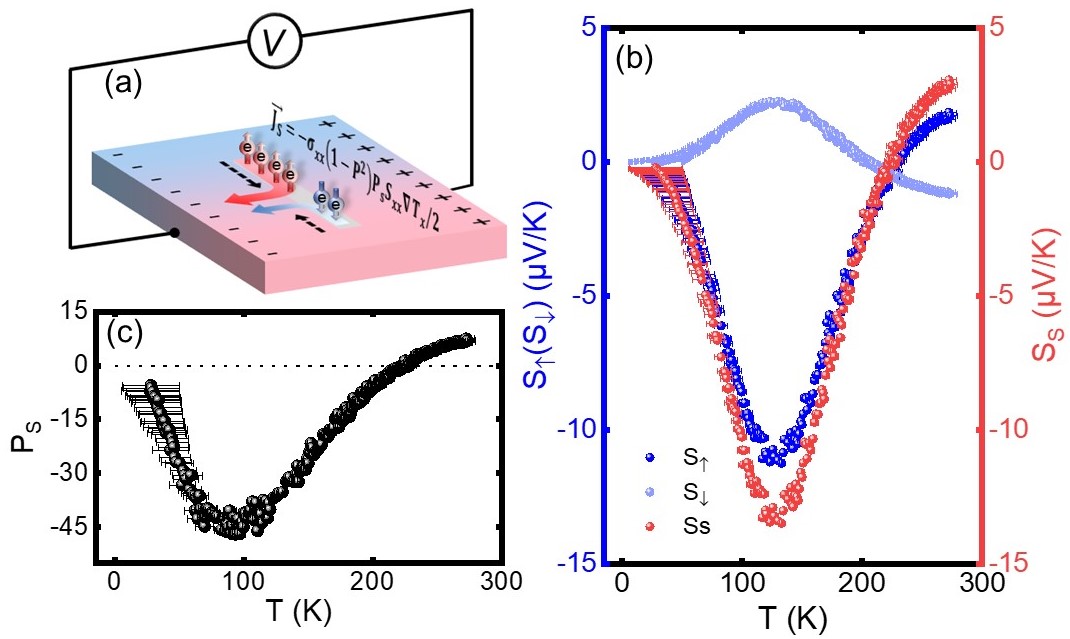}
    \caption{(a) The illustration of ANE generation process based on thermo-spin current. (b) The temperature dependence of Seebeck coefficients for spin-up ($S_{\uparrow}$), spin-down ($S_{\downarrow}$), and spin-dependent Seebeck coefficient ($S_s$). (c) The temperature dependence of thermal spin polarization ($P_s\ =\ S_s/S_{xx}$).} 
    \label{6}
\end{figure}
\end{document}


\title{Supporting Information for Efficient thermo-spin conversion in van der Waals ferromagnet FeGaTe
}

\author{Shuhan Liu}
\affiliation{Department of Physics, Kyushu University, 744 Motooka, Fukuoka, 819-0395, Japan}

\author{Shaojie Hu}
\email[]{hu.shaojie@phys.kyushu-u.ac.jp} 
\affiliation{Department of Physics, Kyushu University, 744 Motooka, Fukuoka, 819-0395, Japan}

\author{Xiaomin Cui}
\affiliation{Department of Physics, Kyushu University, 744 Motooka, Fukuoka, 819-0395, Japan}

\author{Takashi Kimura}
\email[]{t-kimu@phys.kyushu-u.ac.jp}
\affiliation{Department of Physics, Kyushu University, 744 Motooka, Fukuoka, 819-0395, Japan}

\date{\today}

\maketitle


\section{Thickness-dependent Anomalous Hall Effect}
Figure S1 presents the room-temperature AHE curves for three devices with different thicknesses. The main text only focuses on Device\#1 with a thickness of about 75nm via TEM evaluation. The thickness of device\#2 and device\#3 are 65nm and 115nm, respectively. As the black solid curve displays the most substantial AHE signal and a hysteresis loop closer to a rectangle-shape.  The result indicates thinner FGT flakes tend to exhibit higher perpendicular magnetic anisotropy.

\renewcommand{\thefigure}{S\arabic{figure}}
\setcounter{figure}{0}
\begin{figure}[h]
    \centering 
    \includegraphics[width=4in]{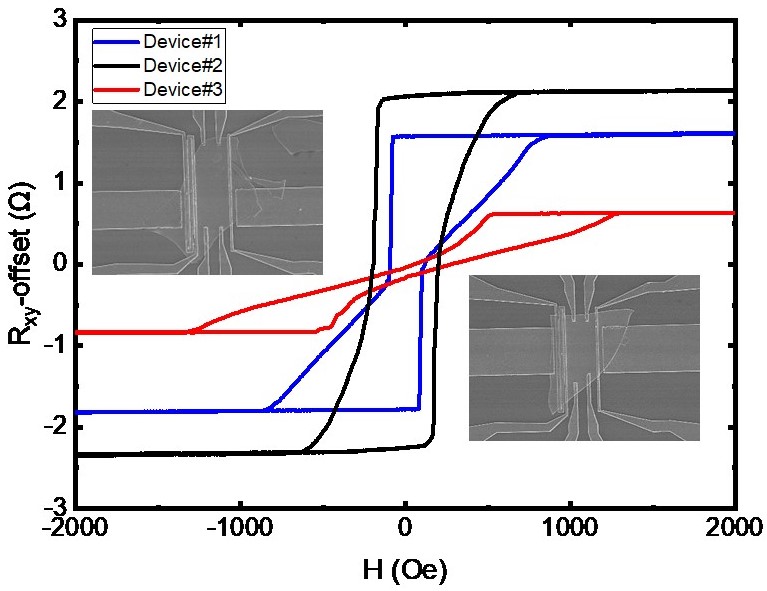}
    \caption{AHE curves of three devices with different thicknesses at 295K. The top and bottom inserts refer to the SEM images of Device\#2 and Device\#3, respectively.}
    \label{S1}
\end{figure}
\section{Temperature gradient calibration}
\begin{figure}[h]
    \centering 
    \includegraphics[width=6in]{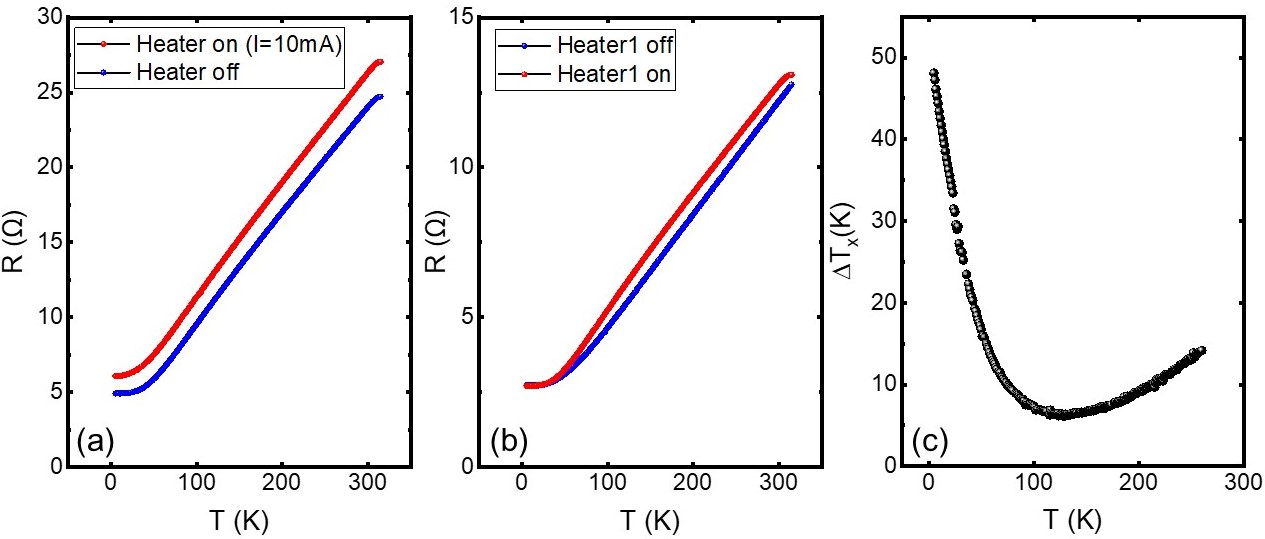}
    \caption{The temperature dependence of resistance for heater sensor 1 (a) and heater sensor 2 (b), and the calculated temperature difference between two heater sensors (c).}
    \label{S2}
\end{figure}
The basic temperature calibration of two heaters was conducted by measuring the temperature dependence of heater resistances in heater-off and heater-on states, respectively. For heater-off state, a very small A.C current was applied to rule out the generated Joule heating effect and make sure the same temperature of two heaters with the cryostat. For heater-on state, a heating current of 10mA was applied in heater 1 to heat one side of the FeGaTe due to the large amount of local Joule heating. Note that 'on' and 'off' here merely refer to the state of H1, while heater2 (H2) only served as a temperature reference of the cold side, which is always applied a small A.C. throughout the whole experiment. Then the temperature difference can be extracted by comparing the resistance changes of two states. Figure S2(c) shows the temperature difference between two heaters ($\Delta T_H$) first decreases and then increases with temperature reduction. The increase below about 100K may be attributed to the phonon-freezing phenomenon, which results in the low thermal transfer efficiency from hot side to cold size, and thus $\Delta T_H$ shows to be greatly promoted\cite{liProximitymagnetizedQuantumSpin2022}. These experimental data were then taken as a reference for COMSOL simulation, which offers further insight into temperature gradient evaluation.

\begin{figure}[h]
    \centering 
    \includegraphics[width=6in]{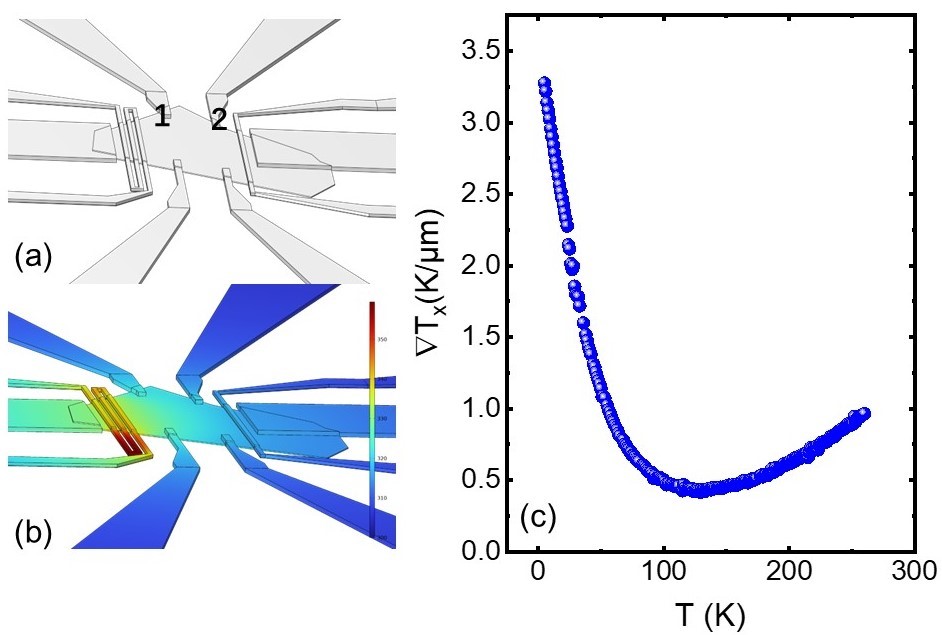}
    \caption{(a) The schematic of simulation structure. (b) The temperature distribution at a base temperature of 260 K. (c) Estimated values of $\nabla T_x$ at different base temperatures.}
    \label{S3}
\end{figure}

We conducted the COMSOL simulations using a model constructed identically in structure and size to our real device, as depicted in Fig.S3(a). 
The physical module of Joule heating was implemented, with the thermal boundary condition fixed at base temperature. The COMSOL-built-in parameters of Si/SiO$_2$ substrate and copper were used, with their thermal conductivities being 1.31 W/(mK) and 317 W/(mK), respectively; while we replaced the unknown parameters of FGT using those of Cu. Such a replacement should be regarded as reasonable, and its impact on the temperature distribution of the whole system should be negligible. This is because as Fang et al. illustrated in their work to investigate ANE of Fe$_3$GeTe$_2$, that the heat transport is dominated by the substrate, where they used Pt to replace Fe$_3$GeTe$_2$ to evaluate the temperature gradient\cite{fangObservationLargeAnomalous2019}. To verify the rationality of this assumption, we first compared the simulation results with experimental data, finding a good alignment between the two. For example, at a base temperature of 260K, the experimental temperature difference between H1 and H2 ($\Delta T_H$) is 14.4K as shown in Fig.S2(c). Our simulation, illustrated in Fig.S3(b), yielded a nearly identical $\Delta T_H$ value of 14.6K. Next, we further explored the temperature difference between electrodes 1 and 2 ($\Delta T_x$), which will be then used for the temperature gradient calibration. $\Delta T_x$ is simulated to be 3.26 K, being about 23.3\% of $\Delta T_H$. By expanding this quantitative relationship to the entire base temperature range, we obtained the values of temperature gradient $\nabla T_x$ at different base temperatures, as shown in Fig.S3(c).

\section{Anomalous Nernst Effect of another device}
The same measurements of ANE were also conducted for Device\#2, as Fig.S4(a) shows. The trends of magnetic field sweeping results of ANE are consistent with those reported in the main text, which suggests the observed sign-reverse behavior of thermal voltage could be attributed to the intrinsic characteristics of FGT itself. Note that the small values of ANE signal originate from the applied small heating current of 3mA, which is significantly lower than that of 10mA used in Device\#1. 

\begin{figure}[h]
    \centering 
    \includegraphics[width=6in]{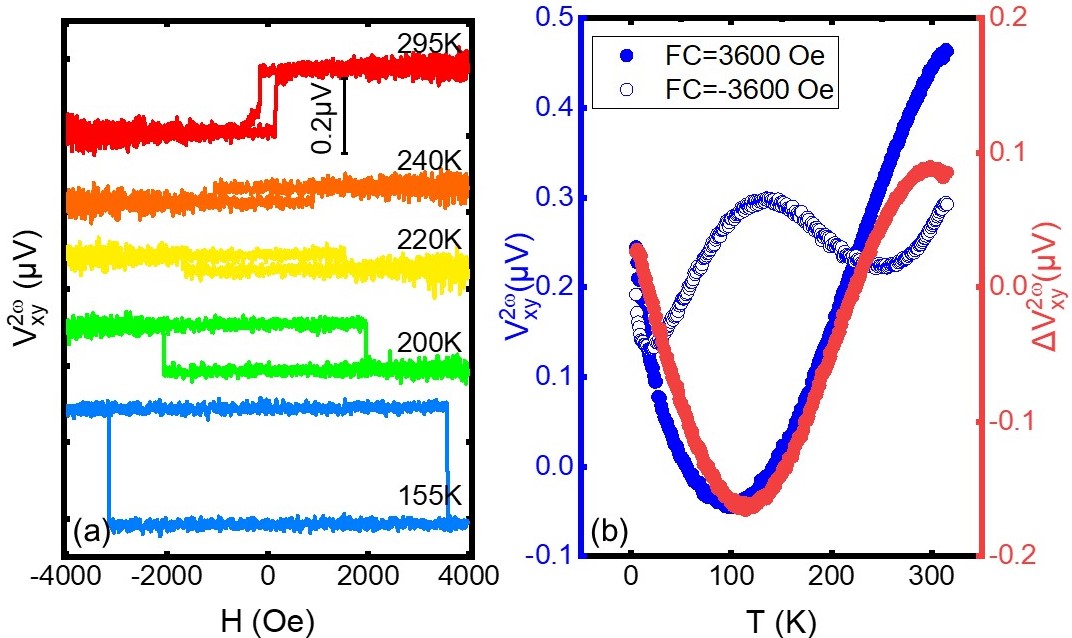}
    \caption{(a) ANE signals obtained by out-of-plane magnetic field sweeping at different temperatures. (b) The temperature dependence of transverse voltage under positive and negative magnetic fields, and the derived ANE voltage. }
    \label{S4}
\end{figure}
\section{Thermo-spin polarization in different materials}

Fig.S5 shows the value range of thermo-spin polarization of CoFeAl\cite{hu2014efficient}, Py\cite{hu2018substantial}, and FeGaTe, which clarifies that $P_S$ in FeGaTe is significantly higher than that of CFA and Py.

\begin{figure}[h]
    \centering 
    \includegraphics[width=4in]{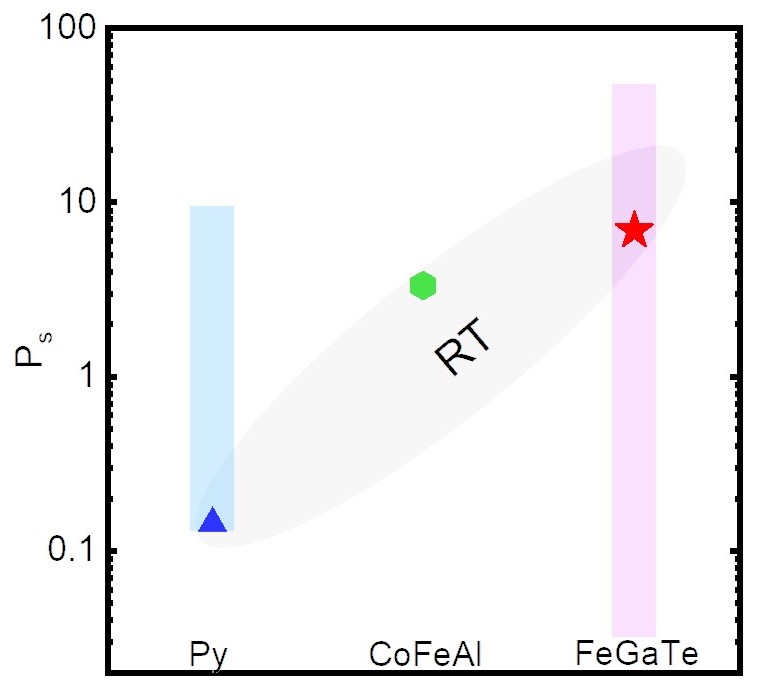}
    \caption{Thermo-spin polarization in different materials. The data within the gray oval region refer to $P_S$ obtained around room temperature}
    \label{S5}
\end{figure}
\newpage
\bibliographystyle{unsrt}
\bibliography{SI}